# On The Peer Review Reports: Does Size Matter?

Abdelghani Maddi[1] and Luis Miotti[2]

[1]*abdelghani.maddi@cnrs.fr*
GEMASS – CNRS – Sorbonne University, 59/61 rue Pouchet 75017 Paris France.

[2] *egidio.miotti@hceres.fr*
Observatoire des sciences et techniques – Hcéres, 2 rue Albert Einstein, 75013 Paris.

**Abstract**

Amidst the ever-expanding realm of scientific production and the proliferation of predatory journals, the focus on peer review remains paramount for scientometricians and sociologists of science. Despite this attention, there is a notable scarcity of empirical investigations into the tangible impact of peer review on publication quality. This study aims to address this gap by conducting a comprehensive analysis of how peer review contributes to the quality of scholarly publications, as measured by the citations they receive.

Utilizing an adjusted dataset comprising 57,482 publications from Publons to Web of Science and employing the Raking Ratio method, our study reveals intriguing insights. Specifically, our findings shed light on a nuanced relationship between the length of reviewer reports and the subsequent citations received by publications. Through a robust regression analysis, we establish that, beginning from 947 words, the length of reviewer reports is significantly associated with an increase in citations.

These results not only confirm the initial hypothesis that longer reports indicate requested improvements, thereby enhancing the quality and visibility of articles, but also underscore the importance of timely and comprehensive reviewer reports. Furthermore, insights from Publons' data suggest that open access to reports can influence reviewer behavior, encouraging more detailed reports.

Beyond the scholarly landscape, our findings prompt a reevaluation of the role of reviewers, emphasizing the need to recognize and value this resource-intensive yet underappreciated activity in institutional evaluations. Additionally, the study sounds a cautionary note regarding the challenges faced by peer review in the context of an increasing volume of submissions, potentially compromising the vigilance of peers in swiftly assessing numerous articles.

**Acknowledgement**

The present paper is an extended version of the contribution presented in the nineteenth conference of the International Society of Scientometrics and Informetrics (ISSI2023) held in Bloomington, Indiana, USA, July 2-5. See: https://zenodo.org/records/8305973

**Compliance with Ethical Standards**

The authors have no relevant financial or non-financial interests to disclose.

**Introduction**

The number of scientific journals has grown exponentially over the past twenty years that it is "*it is now impractical for a researcher to comprehend the entirety of published literature within their field*" (Larivière, 2017). This provides fertile ground for the development of predators seeking to take advantage of the situation. The number of predatory journals is currently estimated at 13,000 journals with more than hundred thousand published articles (Boukacem-Zeghmouri et al., 2020; C. Shen & Björk, 2015).

Boukacem-Zeghmouri et al. (2020) provided a detailed description of the predation phenomenon in the scholarly publishing. Predatory journals are characterized by three main strategies: all predatory journals are Open Access (OA) journals based on a "author-pays" business model, in which peer review is not rigorous (or even absent) and displays "standard markers of scientific publications" (e.g. ISSN, DOI, etc.).

As evidenced in much of the existing literature (Cobey et al., 2019; Siler, 2020), the landscape of scholarly publishing defies simplistic categorization as either black or white. The demarcation between predatory journals and legitimate scientific publications is often blurred, underscoring the need for nuanced evaluation. One crucial criterion for discernment lies in the quality of peer review employed by journals, especially when other administrative attributes associated with publication, such as ISSN and DOI, remain authentic and unaltered. This circumstance is exemplified in the case of recently established journals with limited recognition and modest bibliometric indicators, such as a low impact factor. Faced with the challenges of thriving in the contemporary oligopolistic scholarly communication market, characterized by substantial entry barriers, these journals may resort to adopting assertive commercial strategies to secure growth and viability. Therefore, they may look like predatory journals without actually being.

This issue creates a certain anxiety among the scientific community who fear for their reputation if they happen to publish in journals, which prove to be predatory, or without scientific basis (Frandsen, 2019; Kolata, 2017). Some have described a kind of *omerta* within the scientific community (Djuric, 2015). In this regard, Björk & Solomon (2012) stressed that the OA status of journals should not be a reason for distrust for researchers, since not all OA journals are predatory. Björk and Solomon (2012) have also emphasized that researchers should, on the other hand, carefully check the quality standards of journals before submitting their work.

In this context, the role of peer review is paramount and deserves special attention. How can we "assess" peer review? Does the length of reviewers' reports improve the quality of manuscripts? Alternatively, is it only a rhetorical instrument intended for publishers and authors? The analysis carried out by Publons team is the only one to our knowledge that has investigated the link between the size of reviewers' reports and the citation impact on a large sample of over 378,000 reviews. Publons (2018) showed the existence of a positive correlation between the average number of words in evaluation reports and the impact of journals. Nevertheless, the study concluded that it is not possible to say with certainty that longer reviews are better or worse than shorter ones: "*Great reviews can be short and concise. Poor reviews can be long and in-depth, or vice-versa*". The analysis carried out by Publons remains quite descriptive and does not indeed allow confirming the positive link observed between the two variables. For this, an econometric model is necessary to neutralize the effects of the variables affecting citation impact on the one hand, and the length of the reviewers' reports on the other. In this study, we hypothesize that reviewers' reports actually improve the quality of publications. The interest they have received within the scientific community and results in a high number of citations reflects to some extent their quality. Thus, the length of reviewers' reports is representative of the number of modifications / improvements that authors must make to their manuscript for an eventual acceptance for publication. In other words, there should be a positive relationship between the length of reviewers' reports and the quality of scientific publications.

To test this hypothesis, we use on the one hand the data provided by Publons (https://publons.com/) for the length of reviewer' comments, and on the other hand on the data from the Web of Science (WoS) database for the calculation of bibliometric indicators. We performed an econometric model to analyze the link between the two variables: size of reviewers' reports and citation impact.

It should be noted that the measure used as a proxy for the "quality" of publications, namely citations, is subject to contestation (Aksnes et al., 2019; Leydesdorff et al., 2016). As demonstrated in the literature, citations encompass sociological, strategic, and stochastic dimensions (Waltman, 2016; Wouters, 2020). In the face of the quantitative evaluation of research, citations have become a target, leading to questionable behaviors, including the recourse to self-citation, to manipulate this system strategically (Griesemer, 2020). Furthermore, the literature has documented the sociological dimension of citations, giving rise to what De Solla Price termed "invisible colleges," where groups of authors mutually cite each other, and instances of courtesy citations (Price, 1963). In addition, due to the proliferation of publications, authors must make choices, at times arbitrary, when including articles to cite. This being said, other research has shown that the venue of publication can be a reliable means of assessing research quality, despite occasional instances of misconduct even in esteemed journals such as The Lancet or Nature. As emphasized by Waltman & Traag (2021), reputable journals generally adhere to rigorous editorial processes. And, as highlighted by Traag (2021), prominent high-impact journals strategically choose articles anticipated to garner increased citations, either owing to their relevance to contemporary issues or their significance. For this reason, even though we are cognizant of the multifaceted aspects that may alter the nature of citations, they remain a viable proxy for quality, or at least for the visibility and interest a publication has elicited within the scientific community.

On the side of reviewer reports, one might reasonably posit that the majority of reports requiring extensive work and corrections tend to be of a certain length. Conversely, reports necessitating minimal modifications are generally more concise. While we acknowledge that exceptions to this general rule may exist, with succinct reports substantially altering papers and lengthy reports making only marginal modifications, these instances are recognized as outliers.

While recognizing the potential influence of variability in review content expectations on report length, it is crucial to underscore that our study primarily focuses on investigating the impact of report length, irrespective of its origin—whether intrinsic to reviewers or influenced by journal guidelines or authors' encouragement. Although different journals may shape review content expectations differently, our analysis remains robust in exploring the association between the length of reviewer reports and the subsequent citation impact on publications. Our study aims to provide a comprehensive overview of this relationship, acknowledging the broader context of diverse review structures without compromising the validity of our core findings.

The article is structured as follows: Initially, we present a literature review exploring the association between peer review, in a general context, and the quality of scholarly publications. Following this, descriptive statistics derived from the employed database (Publons) are provided. Subsequently, the methodological approach, particularly the regression analysis utilized in our study, is introduced. This leads into the results section. Finally, the conclusion and discussion sections revisit the obtained results, elucidating their broader implications and significance within the framework of our study.

**Literature review**

Peer review is a topic that has garnered much ink in the literature in Scientometrics and the sociology of science. We will only present here the work related to our research question, which is the impact of the peer review process on the quality of publications, approached by citations indicators.

Peer review processes are generally long, with disparities across disciplines (Björk & Solomon, 2013; Cornelius, 2012; Huisman & Smits, 2017; Kareiva et al., 2002). Especially for accepted publications (Bilalli et al., 2020). Over the past ten years, the duration has lengthened because of the inflation in the number of publications submitted to journals. On the other hand, the acceptance rate has increased by around 50% (Björk, 2018). In addition to the constraints related to the number of publications submitted each week to journals, the fact that referees are not paid for their review activity can contribute to slowing down the process (Azar, 2007; Moizer, 2009; Toroser D et al., 2016). The increase in the technicality of publications can also increase the evaluation time in certain disciplines (Azar, 2007).

Whatever the reasons that may affect the evaluation times of scientific publications, the final objectives of this are multiple and revolve around improving the effective quality of publications. (Chataway & Severin, 2020) outlined eight purposes of peer review: (1) assess the contribution and originality of a manuscript, (2) perform quality control, (3) improve manuscripts, (4) assess the suitability of manuscripts for the topics of the journal, (5) provide a decision-making tool for editors, (6) provide comments and peer feedback, (7) strengthen the organization from the scientific community, and (8) provide some sort of accreditation for published papers.

Studies that assess the quality of the review process, from the perspective of authors or editors, are quite rare. Huisman & Smits (2017) analyzed data from 3,500 experiences and author reviews published on SciRev.sc (https://scirev.org/). The SciRev.sc interface groups together author reviews and ratings assigned to journals based on the quality (perceived by them) of the review process. Huisman & Smits (2017) have shown that, unsurprisingly, reviews with short response times tend to score higher. The same goes for experiences that resulted in a positive response from the journal. Drvenica et al. (2019) arrived at similar results on a sample of 193 authors. Furthermore, (Huisman & Smits, 2017) have shown that journals in disciplines where the evaluation processes are relatively long are on average better rated than journals in disciplines where the processes are short. In a more recent study, Pranić et al. (2020) analyzed the perception of authors and editors of the quality of peer reviews in 12 journals: 809 manuscripts and 313 opinions and recommendations. Pranić et al. (2020) found that authors give high ratings and positive perception when reviewers' comments recommend acceptance, unlike comments that recommend rejection (which is to be expected). On the other hand, the evaluations recommending a revision, are of better quality according to the indicator used (*Review Quality Instrument* - RQI). In addition, Pranić et al. (2020) have shown a strong association between the recommendation of referees and the publication decision of editors.

Research that investigates the duration of the peer review process along with the impact of journals or publications is notably infrequent. The study of Pautasso & Schäfer (2010) on 22 ecological / interdisciplinary journals, showed the existence of an inverse relationship between the acceptance delay and the impact factor of the journals. These same journals, however, have relatively low acceptance rates. The analysis of (S. Shen et al. (2015) on the three journals (Nature, Science and Cell), over the period 2005-09, arrived at different conclusions from those of Pautasso & Schäfer (2010). For the three journals studied, the authors observed an inverse relationship between editorial time and the number of citations received.

There is room for improvement in the existing literature on assessing the length of the peer review process. The analyzes of Huisman & Smits (2017), of Drvenica et al. (2019) and of Pranić et al. (2020) can be criticized insofar as they assess the perceived quality of the evaluation and not the intrinsic quality. The perceived quality depends very much on the final decision and the length of the revisions requested by the referees; it is therefore very subjective. Hence, a thorough analysis of referees can result in a lot of criticism and be a source of improvement of the publications, even if the authors perceive it negatively, because it delays the publication (and adds a lot of additional work). To measure intrinsic quality, it is essential to cross-reference the performance indicators (eg citations received) of journals or articles.

Likewise, the study by Pautasso & Schäfer (2010) which showed that high impact journals have short turnaround times ignore the type of decision made by the journals. In general, the number of articles submitted to high-impact journals is important, as is the rejection rate. The rejection decision usually comes a few days / weeks after submission, reducing the average processing time for the entire review. Taking into account the time to get the first response and the type of response are important elements in the analysis. Thus, it can be assumed that lead times for accepted papers are higher in high-impact journals.

**Data**

*Web of Science data*

The data about citations scores and disciplinary assignation of publications has been extracted from the "Observatoire des Sciences et Techniques" (OST) in-house database. It includes five indexes of WoS available from Clarivate Analytics (SCIE, SSCI, AHCI, CPCI-SSH and CPCI-S. for more information see: https://clarivate.com/webofsciencegroup/solutions/webofscience-platform/) and corresponds to WoS content indexed through the end of November 2020. We have limited the analysis only to the original contributions; i.e. the following documents types: "Article", "Conference proceedings" and "Review".

*Publons data*

Publons is an interface created in 2012 specializing in peer review issues. In 2017, the supplier of the WoS database, *Clarivate Analytics*, acquired Publons. Currently, this database indexes over 2 million researcher profiles who share their peer review experiences. Publons offers a free service for researchers to promote their contributions. Publons also made available more than 300,000 items designating the characteristics of referee evaluation reports in journals: average word count, country of referee and journal. As part of this study, Publons made its database available to us by adding article identifiers (WoS UT). This allowed us to easily match Publons data with that of the WoS database and add other variables. Table 1 shows the distribution of publications according to the number of reports available in Publons.

**Table 1: Number of reports per publication**

| #Reports by publication | #Publications | % | #Reports | % |
|---|---|---|---|---|
| 1 | 57 256 | 93.56% | 57 256 | 87.54% |
| 2 | 3 691 | 6.03% | 7 382 | 11.29% |
| 3 | 233 | 0.38% | 699 | 1.07% |
| 4 | 16 | 0.03% | 64 | 0.10% |
| 5 | 1 | 0.00% | 5 | 0.01% |
| **Total** | 61 197 | 100.00% | 65 406 | 100.00% |

*Data preparation and preprocessing*

The data provided by Publons offer, for each publication (UT WoS), the word count in the reviewers' reports. Unfortunately, information regarding the peer review stage (1st round, 2nd round, etc.) is not available. Additionally, only the word count of a single report is available for the majority of publications. Due to the random nature of this information (i.e., the variable number of available reports per publication), we conducted a random selection of one report per publication for those with multiple reports, ensuring a singular report per publication. This resulted in a total of 61,197 distinct publications. This solution seems optimal because the Publons database does not provide the number of reviewers (number of reports) by document and, consequently, the use of procedures such as the clustering of residues seems to be not necessary in this case.

As we intend to conduct regression analysis, with one of the variables being the number of funders per publication, we limited our dataset to publications with a publication year after 2009. This choice was made because funding information was not well-documented in the metadata before this date. Furthermore, we restricted the dataset to publications adhering to OST criteria, such as being citable, having complete metadata, and specific document types (articles and reviews). Following this filtering process, the sample size was reduced to 57,694 publications.

These data underwent preprocessing to ensure the quality and robustness of the analysis results. We computed the Interquartile Range (IQR) of the variable "length of reports," expressed in terms of word count, to assess data dispersion. An exclusion threshold for extreme values was defined by multiplying the IQR by a predetermined factor (equal to 5), and observations outside this range were excluded (up to a maximum of 13,671 words for a single report). This step was crucial to prevent outlier values from unduly influencing our regression model. Consequently, this procedure led to the exclusion of 212 extreme observations. Therefore, the sample used for the initial regression consists of 57,482 publications.

Table 2 provides the distribution of the number of publications by reviewer report length interval.

**Table 2: descriptive statistics for intervals (review length words)**

| Lower bound | Upper bound | Number | % |
|---|---|---|---|
| *0* | *200* | 21 007 | 37.0% |
| **200** | **400** | 13 946 | 24.0% |
| **400** | **600** | 9 052 | 16.0% |
| **600** | **800** | 5 252 | 9.1% |
| **800** | **1000** | 3 114 | 5.4% |
| **1000** | **1200** | 2 001 | 3.5% |
| **1200** | **1400** | 1 111 | 1.9% |
| **1400** | **1600** | 706 | 1.2% |
| **1600** | **1800** | 418 | 0.7% |
| **1800** | **2000** | 303 | 0.5% |
| **2000** | **2200** | 220 | 0.4% |
| **2200** | **2400** | 147 | 0.3% |
| **2400** | **2600** | 104 | 0.2% |
| **2600** | **2883** | 101 | 0.2% |
| **Total** | | **57 482** | **100.0%** |

| | |
|---|---|
| *Number of observations* | 57 482 |
| *Average* | 416.3 |
| *Median* | 302 |

As we can see from the table 2, the largest category of reports, approximately 37.0%, fall within the range of 0 to 200 words. The second range, from 200 to 400 words, accounts for 24.0% of the total, closely followed by the 400 to 600 words range, representing 16.0%. Analyzing the aggregated statistics, the average number of words per peer review report is 416.3, while the median is 302. These values highlight some dispersion in the distribution, indicating the presence of significantly longer reports that influence the average.

**Method**

*Adjustment of Publons data*

The data extracted from Publons database represents a sample of the global WoS database, which we consider in this work as the population. When working with samples (e.g. public opinion polls), it is important to adjust data for population parameters. Thus, to study whether the conclusions drawn on the Publons database can be extended to the WoS database, we analyzed whether the respective structures are similar, based on a certain number of variables considered significant in explaining citation scores. Given that there are statistically significant differences between the Publons structure and that of the WoS, we adjusted the Publons sample using the procedure called "raking ratio" (Deming & Stephan, 1940; Deville et al., 1993), which provides adjustment weights. The variables used for the adjustment of the Publons sample:
- Publication year: dummy variables;
- Open access: binary variable (Yes/No);
- Financing: one, two, three, four or more;
- Number of countries: one, two, three, four or more;
- Scientific discipline: ERC codes (dummy variable).
- The impact of journal (5 classes : <0.8, [0.8 , 1.2[, [1.2 , 1.8[, [1.8 , 2.2[, >=2.2), for the calculation method see: (Maddi & Sapinho, 2022).

Figure 1 summarizes the Publons data adjustment procedure. Appendix A.1 and A.2 show the structure of the Publons sample and of the relevant WoS population, before and after adjustment.

**Figure 1: Adjustment of Publons data and regression methods**

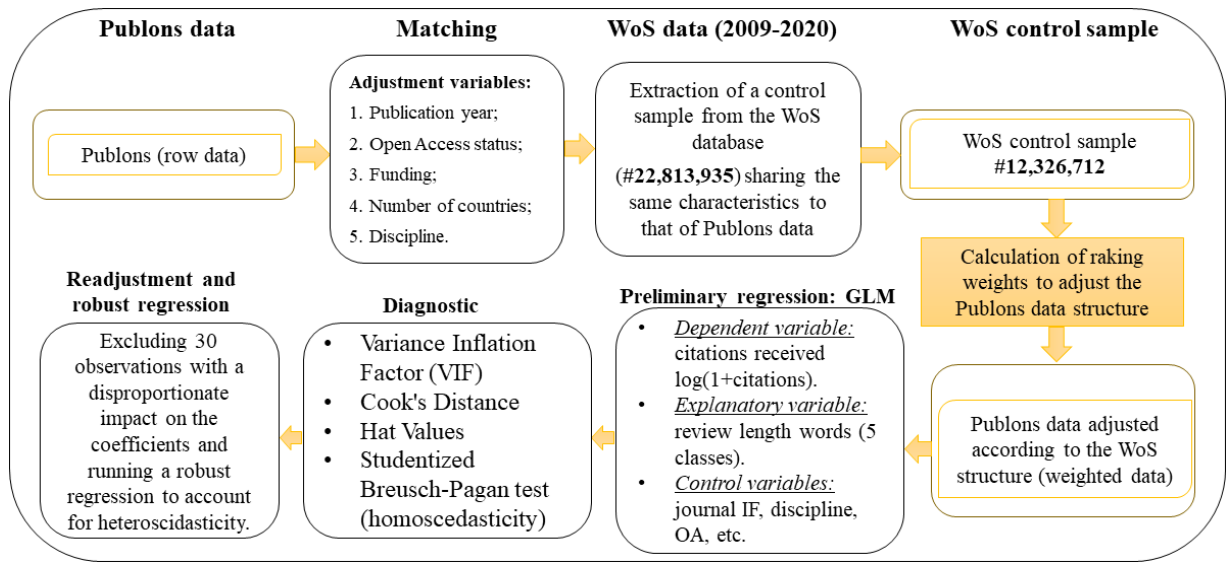

*Regression analysis*

In their 2014 study, Thelwall and Wilson explore the escalating use of citations in research assessment. The authors emphasized the crucial need to identify factors influencing citation scores independently of scholarly quality, advocating for the application of regression as the most potent statistical technique in this context.

The study's findings reveal that citation counts typically adhere to a discrete lognormal distribution, in contrast to previous investigations employing negative binomial regression. Thelwall & Wilson (2014) propose an alternative approach involving the addition of one to each citation, logarithmic transformation, and subsequent utilization of the general linear model (ordinary least squares) for regression, or the use of the generalized linear model without logarithmic transformation. Their results, based on simulated discrete lognormal data, underscore the effectiveness of this strategy. Accordingly, in alignment with the guidelines proposed by Thelwall & Wilson (2014), we have chosen to employ the Generalized Linear Model (GLM) as the favored approach for citation modeling. Our response variable consists of log-transformed citations ($\log(1 + \text{number of citations})$), incorporating an addition of 1 prior to transformation to address cases where the citation count is zero.

The principal independent variable in our model is the word count in reviewers' reports. To enhance its suitability for analysis, we discretized this variable into distinct categories. The category intervals were determined based on the data distribution characteristics, utilizing the Fisher discretization method. This process resulted in the identification of five discrete classes of reviewers' reports, as detailed in Table 3.

**Table 3: Discretization of the report length variable into 5 classes**

| Classes | N = **57 482**[1] |
|---|---|
| Less than 232 words | 23 532 (41%) |
| 232 to 535 words | 18 041 (31%) |
| 536 to 946 words | 10 123 (18%) |
| 947 to 1 612 words | 4 532 (7.9%) |
| 1 613 to 2 891 words | 1 254 (2.2%) |

[1] n (%)

In addition to the primary variable, several control variables were included in our model. The selection of these control variables drew from the literature, considering their potential impact on received citations. These variables encompass the journal's impact factor, the number of countries involved in the publication (indicative of international collaboration) (Ni & An, 2018), the number of funders (Quemener et al., 2023; Yan et al., 2018), the publication's open-access status (Maddi & Sapinho, 2023), the discipline of the publication (Lillquist & Green, 2010), and the publication year (Bornmann, 2013). These variables were integrated into the model to capture and control for potential effects on our dependent variable, citations. Table 4 provides a detailed description of the dependent, independent, and control variables

**Table 4: Model variables by type**

| Type | Variable | Designation |
|---|---|---|
| **Dependent variable** | Citations | We utilized the logarithm of (1 + the total number of citations) as the dependent variable. |
| **Explanatory variable** | Review length (words) | The independent variable in our study is the word count of reviewers' reports, categorized into discrete intervals using the Fisher discretization method. |
| **Control variables** | Journal impact factor | The classic two-year impact factor. |
| | Open Access | The status of open access: Yes / No (with "No" as a reference value). |
| | Funding | The logarithm-transformed variable represents the number of grants received, derived from the WoS acknowledgments field, ranging from 0 (indicating no funding) to n. |
| | International collaboration | The logarithm-transformed variable represents the number of countries involved in the publication (using author addresses), ranging from 1 (publications without international collaboration) to n. |
| | Discipline | The discipline variable encompasses the 14 fields provided in the Publons dataset. While we used the more detailed 27+2 discipline nomenclature from the OST (https://figshare.com/articles/dataset/OST_-_Classification_of_WoS_subject_categories_into_27_2_ERC_panels_/21707543) for fine-tuning, the regression analysis was streamlined by utilizing the 14 disciplines as effective control variables. |
| | Publication year | The publication years, between 2010 and 2020 |

*Diagnostic of preliminary regression and model readjustment*

After the initial model adjustment, diagnostic assessments were conducted to evaluate the robustness of the results. These diagnostics included residual plots, influence tests, and a multicollinearity analysis employing the Variance Inflation Factor (VIF) (see appendixes A3 and A4). The objective is to enhance the credibility of the findings by identifying potential issues related to collinearity among the explanatory variables.

With regard to the VIF tests, all values are significantly below 5 (around 1), suggesting the absence of multicollinearity. Cook's Distance and Hat Values were employed as measures of influence. These metrics facilitated the identification of potentially influential observations, thereby determining those with a disproportionate impact on the model coefficients. Thresholds used for Cook's Distance and Hat Values were set at 0.02 and 0.01, respectively. To preserve the integrity of the model, we chose to exclude observations with a disproportionate impact, totaling 30 observations.

The exclusion of influential observations was followed by a subsequent model readjustment. The readjusted model underwent new diagnostic assessments, including residual plots and tests for homoscedasticity (studentized Breusch-Pagan test).

Upon detecting heteroscedasticity, a robust regression with robust standard errors was adopted. This approach seeks to provide robust estimates even in the presence of potential violations of classical linear regression assumptions. The robust approach involves calculating standard errors that do not rely on specific assumptions about the distribution of errors. Unlike ordinary least squares, this approach offers a more resilient alternative in the face of violations of classical conditions, such as heteroscedasticity.

**Results**

The analysis of the regression results (figure 2) sheds light on significant relationships between various variables and the number of citations received by publications. Particularly, the length of peer review reports appears to play an important role, as evidenced by the estimated coefficients for different length categories.

Examining the coefficients for report length intervals, it is noteworthy that the [947,1612] and [1613,2891] categories exhibit positive and significant coefficients. This suggests that peer review reports with a length between 947 and 2891 words are associated with a substantial increase in the number of citations received by a publication. This observation aligns with the initial hypothesis that longer reports may be correlated with improved article quality, translating into greater visibility and academic impact.

Other control variables also display significant coefficients, underscoring their importance in predicting the number of citations. Factors such as the number of countries involved in the publication, the number of funders, open access status, publication year, journal impact factor, and publication discipline all exhibit significant effects on the number of citations received.

Crucially, these results stem from a robust analysis, accounting for factors such as influential observations, and heteroscedastic errors. This enhances confidence in the validity of the conclusions drawn from the regression. In summary, these findings suggest that the length of peer review reports can significantly influence the visibility and impact of publications.

**Figure 2: Results of estimates of the impact of the length of reviewers' reports on citations**

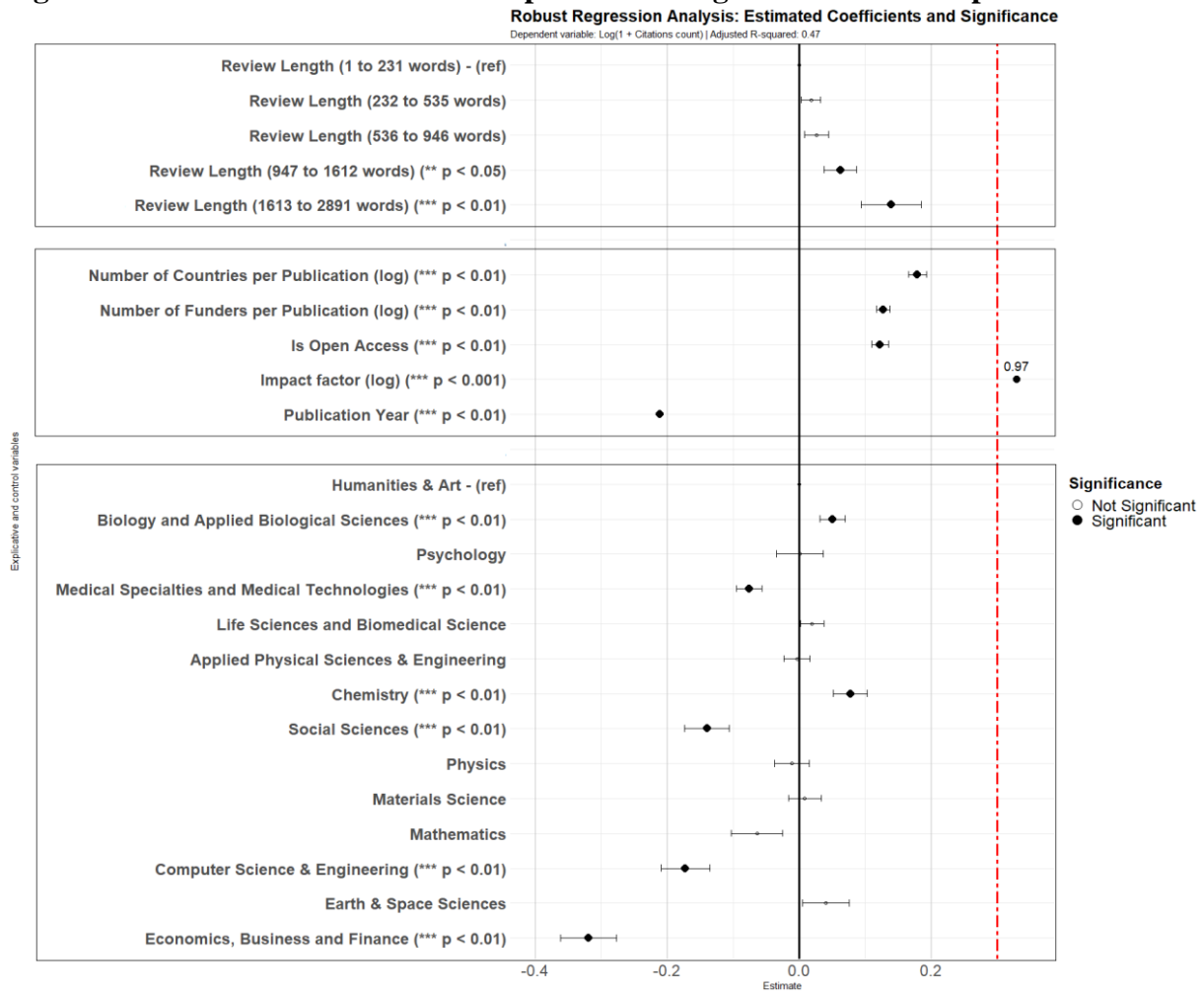

However, we should take this result with some caution for several reasons. If the correction of the structural bias allows a comparison of the Publons database with the control sample taken from WoS, it does not totally eliminate the selection bias. This selection bias appears if the publication of the reviewers' reports in Publons is not done randomly but meets specific criteria (personal -age, gender-, and professionals -seniority, affiliation, scientific profile-, journal constraints concerning obligation/choice to publish reports, etc.). A few leads suggest the existence of a selection bias:

- The Open Access rate in Publons is 39.1% (and 43% in the case of a sample limited to reports with a word count in the 200-2891 range) and in the control sample drawn of WoS is 28.7%. This over-representation in Publons may indicate non-random behavior in reviewer reporting. Therefore, it is not impossible that reviewers, who dump their entire reports in Publons, can do so strategically. In other words, we only put on Publons well-produced reports. A recent paper (Radzvilas et al., 2022), based on evolutionary game theory, indicates that the type of peer review of the journal directly affects the behavior of researchers and reviewers. Specifically, the paper indicates that if the journal practices open peer review, it encourages researchers and reviewers to put in more effort (because it directly affects their reputation). Our results seem to indicate this phenomenon (especially on the side of the reviewers) and merit further investigation. Another paper (Bravo et al., 2019) shows that the publication of reports can change the tone used by reviewers (especially younger ones) due in some cases to fears of reprisals.
- The funding rate in Publons is 73.7% (and 76.3% in the case of a sample limited to reports with word counts in the 200-3000 range) and in the WOS control sample is of 53.9%. We know that the impact of funding on the citations of articles is positive and significant and, therefore, to hypothesize better quality. This could prompt some reviewers to declare their reports to Publons.

In sum, our regression results are both interesting and intriguing with respect to the Publons data, as they leave many questions unanswered. To be able to generalize our conclusions, it would be interesting to do the same work on a random sample of peer review reports. One idea would be to carry out the analysis using a panel of journals that practice open peer review, while respecting the problem of representativeness of the data and their structure in relation to the population (WoS, Scopus or any other large database - e.g. OpenAlex).

At this stage, we can simply say that the preliminary results suggest the existence of a positive link between the length of the peer evaluation reports and the quality of the publications (approximated by the citations).

**Discussion and Conclusion**

Through this paper, we have sought to analyze empirically to what extent peer review improves the quality of publications. To do this, we used data from Publons (#57,482) with a structure adjusted to that of the WoS database (control group of 12,326,712 publications). The proxy used for peer review is the length of review reports expressed in number of words (provided by Publons). On the publications side, the proxy used to measure the quality (interest aroused within the scientific community) of the publications are the citations received. We carried out several regression models, each time putting the citation scores as the explained variable, and the length of the reviewers' reports as the explanatory variable (while adding a certain number of control variables). In addition to the "classic" results on the importance of Open Access, funding or international collaboration on citations, we have shown the existence of a strong and statistically significant relationship between the length of reviewers' reports and citations received.

Our exploration into the intricate relationship between the length of peer review reports and the subsequent citation impact of publications has illuminated intriguing facets of scholarly communication. The findings reveal a statistically significant association between the length of reviewer reports, particularly those exceeding 947 words, and a substantial increase in the citations garnered by the published works. This aligns coherently with our initial hypothesis, suggesting that more extensive reports may signal constructive feedback and a commitment to elevating the overall quality of the articles, thereby amplifying their visibility and impact.

The implications of these discoveries extend beyond the statistical intricacies of regression analysis. They underscore the pivotal role of timely and comprehensive reviewer assessments in shaping the scholarly landscape. The observed correlation between longer reports and heightened citations implies that affording reviewers sufficient time for thorough evaluations may be instrumental in enhancing the quality and impact of scholarly publications.

Moreover, our results advocate for the promotion of transparency in the peer review process, exemplified by initiatives like open access to review reports. Such transparency may influence reviewer behavior, fostering more detailed and insightful assessments. This resonates with the findings of studies such as (Radzvilas et al., 2022), highlighting the potential cascading effect of open peer review practices on reviewer engagement and diligence.

The role of reviewers emerges as a central focus. The study emphasizes the imperative to acknowledge and incentivize the often arduous and underappreciated role of reviewers in institutional evaluations. As gatekeepers of scholarly integrity, reviewers play a crucial role in safeguarding against questionable practices and errors that may compromise the quality of published research. However, challenges faced by peer review, exacerbated by the surge in submissions and reviewer demands, warrant careful consideration. The escalating burden on reviewers to assess numerous articles promptly raises concerns about the effectiveness of the evaluation process. It prompts reflection on how to maintain vigilance and uphold rigorous standards in the face of heightened workload and time constraints. The rise of post-publication peer review (PPPR) is a noteworthy development in this context. The increasing prevalence of PPPR suggests a nuanced shift in the traditional peer review paradigm, indicating an evolving landscape where ongoing evaluation and scrutiny persist after publication. This trend raises questions about whether the robust emergence of PPPR is a consequence of the limitations of classical peer review in preemptively identifying issues.

In conclusion, our study offers more than a mere glimpse into the intricate interplay between peer review practices and publication impact. It advocates for a holistic understanding of the peer review process, recognizing its multifaceted role in shaping the scholarly narrative. The correlation between lengthy reviewer reports and increased citations calls for a nuanced approach to reviewer engagement, transparent practices, and ongoing vigilance to maintain the robustness of scholarly communication in an ever-evolving academic landscape.

**Limitations and Future Research**

Despite efforts to rectify the data structure in the Publons database, certain limitations persist. The database relies on researchers voluntarily showcasing their reviewing activities, introducing a potential bias in the dataset's representativeness. Consequently, the findings may not generalize universally due to this selection bias. To mitigate this limitation, it would be valuable to replicate this study with alternative datasets, such as those from journals employing open peer review practices. This approach would involve examining whether similar conclusions (or lack thereof) emerge when utilizing more systematic and randomly sourced data, extracting both publication reports and corresponding citation data.

Another inherent limitation lies in the exclusive reliance on reports from accepted publications. To offer a more comprehensive view, it would be insightful to incorporate an analysis of reviewer report lengths for rejected publications. This extension would provide a nuanced understanding of how the length of reviewer reports may vary across different publication outcomes.

Furthermore, exploring the potential endogeneity concerning the length of reviewer reports is a crucial avenue for further investigation. The concern arises from the possibility that lengthy reports might be indicative of papers already possessing inherent quality. Addressing this issue could involve implementing author disambiguation techniques to control for report length in relation to authors' notoriety. By disentangling the influence of author reputation, we could gain a more nuanced understanding of whether report length genuinely correlates with paper quality or is influenced by external factors. However, it is crucial to acknowledge that long reports may also pertain to submissions of lower quality that ultimately transform into high-quality publications, a nuance our results seem to confirm. This highlights the complexity of peer review and the dynamic transformation of academic works throughout the publication process.

Finally, it would be worthwhile to consider an intriguing avenue for future research, namely, exploring the potential association between the length of reviewer reports and the extent of modifications made to published papers. This could involve a comparative analysis, for instance, between published versions and preprints, shedding light on whether longer reports are linked to more substantial alterations in the publication process. This novel angle could provide valuable insights into the dynamics of peer review and the impact of reviewer feedback on the evolution of scholarly works.

**Appendices**

**A.1. Structure of the Publons sample and of the WoS population, before and after adjustment**

| Variable | Modalities | Number | % | Marginal sums on population | Marginal sums on population (%) | Publons structure after adjustement (%) |
|---|---|---|---|---|---|---|
| Publication year | 2009 | 761 | 1.31 | 552615 | 4.48 | 4.48 |
| | 2010 | 946 | 1.63 | 546 703 | 4.44 | 4.44 |
| | 2011 | 1356 | 2.33 | 717 698 | 5.82 | 5.82 |
| | 2012 | 1762 | 3.03 | 839 311 | 6.81 | 6.81 |
| | 2013 | 2601 | 4.48 | 1002514 | 8.13 | 8.13 |
| | 2014 | 3 521 | 6.06 | 1141004 | 9.26 | 9.26 |
| | 2015 | 4742 | 8.16 | 1 249 900 | 10.14 | 10.14 |
| | 2016 | 5 917 | 10.19 | 1 348 223 | 10.94 | 10.94 |
| | 2017 | 14 336 | 24.68 | 1 591 868 | 12.91 | 12.91 |
| | 2018 | 18 699 | 32.19 | 1679518 | 13.63 | 13.63 |
| | 2019 | 3046 | 5.24 | 1 140 289 | 9.25 | 9.25 |
| | 2020 | 401 | 0.69 | 516787 | 4.19 | 4.19 |
| | 2021 | 5 | 0.01 | 282 | 0.00 | 0.00 |
| Open access | 1 | 25 177 | 43.34 | 3532995 | 28.66 | 28.66 |
| | 2 | 32 916 | 56.66 | 8793717 | 71.34 | 71.34 |
| Without financing | 1 | 15 284 | 26.31 | 5681665 | 46.09 | 46.09 |
| | 2 | 42 809 | 73.69 | 6645047 | 53.91 | 53.91 |
| One financing | 1 | 13 809 | 23.77 | 2613911 | 21.21 | 21.21 |
| | 2 | 44 284 | 76.23 | 9 712 801 | 78.79 | 78.79 |
| Two financing | 1 | 10 758 | 18.52 | 1762755 | 14.30 | 14.30 |
| | 2 | 47 335 | 81.48 | 10563957 | 85.70 | 85.70 |
| Three financing | 1 | 7155 | 12.32 | 921 952 | 7.48 | 7.48 |
| | 2 | 50 938 | 87.68 | 11404760 | 92.52 | 92.52 |
| Four financing or more | 1 | 11087 | 19.08 | 1 346 429 | 10.92 | 10.92 |
| | 2 | 47 006 | 80.92 | 10980283 | 89.08 | 89.08 |
| N Countries | 1 | 41008 | 70.59 | 10640214 | 86.32 | 86.32 |
| | 2 | 12 778 | 22.00 | 1485416 | 12.05 | 12.05 |
| | 3 | 3 336 | 5.74 | 167 730 | 1.36 | 1.36 |
| | 4 | 971 | 1.67 | 33 352 | 0.27 | 0.27 |
| **Total** | | 58 093 | 100.0 | 12 326 712 | 100.0 | 100.0 |

**A.2. Structure of the Publons sample and of the WoS population, before and after adjustment***

| Variable - ERC Class | Modalities | Number | % | Marginal sums on population | Marginal sums on population (%) | Publons' structure after adjustement (%) |
|---|---|---|---|---|---|---|
| LS09 | 1 | 53 | 0.09 | 9 830 | 0.08 | 0.08 |
|  | 2 | 58 040 | 99.91 | 12 316 882 | 99.92 | 99.92 |
| LS1 | 1 | 6 255 | 10.77 | 695 214 | 5.64 | 5.64 |
|  | 2 | 51 838 | 89.23 | 11 631 498 | 94.36 | 94.36 |
| LS2 | 1 | 6 054 | 10.42 | 489 234 | 3.97 | 3.97 |
|  | 2 | 52 039 | 89.58 | 11 837 478 | 96.03 | 96.03 |
| LS3 | 1 | 1 855 | 3.19 | 129 674 | 1.05 | 1.05 |
|  | 2 | 56 238 | 96.81 | 12 197 038 | 98.95 | 98.95 |
| LS4 | 1 | 9 684 | 16.67 | 1 668 900 | 13.54 | 13.54 |
|  | 2 | 48 409 | 83.33 | 10 657 812 | 86.46 | 86.46 |
| LS5 | 1 | 4 350 | 7.49 | 555 326 | 4.51 | 4.51 |
|  | 2 | 53 743 | 92.51 | 11 771 386 | 95.49 | 95.49 |
| LS6 | 1 | 3 805 | 6.55 | 410 871 | 3.33 | 3.33 |
|  | 2 | 54 288 | 93.45 | 11 915 841 | 96.67 | 96.67 |
| LS7 | 1 | 17 372 | 29.90 | 2 824 978 | 22.92 | 22.92 |
|  | 2 | 40 721 | 70.10 | 9 501 734 | 77.08 | 77.08 |
| LS8 | 1 | 8 051 | 13.86 | 770 001 | 6.25 | 6.25 |
|  | 2 | 50 042 | 86.14 | 11 556 711 | 93.75 | 93.75 |
| LS9 | 1 | 8 065 | 13.88 | 1 202 239 | 9.75 | 9.75 |
|  | 2 | 50 028 | 86.12 | 11 124 473 | 90.25 | 90.25 |
| PE09 | 1 | 53 | 0.09 | 9 830 | 0.08 | 0.08 |
|  | 2 | 58 040 | 99.91 | 12 316 882 | 99.92 | 99.92 |
| PE1 | 1 | 2 029 | 3.49 | 326 962 | 2.65 | 2.65 |
|  | 2 | 56 064 | 96.51 | 11 999 750 | 97.35 | 97.35 |
| PE10 | 1 | 7 293 | 12.55 | 1 411 496 | 11.45 | 11.45 |
|  | 2 | 50 800 | 87.45 | 10 915 216 | 88.55 | 88.55 |
| PE11 | 1 | 5 288 | 9.10 | 1 227 477 | 9.96 | 9.96 |
|  | 2 | 52 805 | 90.90 | 11 099 235 | 90.04 | 90.04 |
| PE2 | 1 | 5 452 | 9.38 | 1 039 190 | 8.43 | 8.43 |
|  | 2 | 52 641 | 90.62 | 11 287 522 | 91.57 | 91.57 |
| PE3 | 1 | 3 419 | 5.89 | 659 856 | 5.35 | 5.35 |
|  | 2 | 54 674 | 94.11 | 11 666 856 | 94.65 | 94.65 |
| PE4 | 1 | 6 763 | 11.64 | 1 441 848 | 11.70 | 11.70 |
|  | 2 | 51 330 | 88.36 | 10 884 864 | 88.30 | 88.30 |
| PE5 | 1 | 5 917 | 10.19 | 1 294 056 | 10.50 | 10.50 |
|  | 2 | 52 176 | 89.81 | 11 032 656 | 89.50 | 89.50 |
| PE6 | 1 | 2 039 | 3.51 | 959 256 | 7.78 | 7.78 |
|  | 2 | 56 054 | 96.49 | 11 367 456 | 92.22 | 92.22 |
| PE7 | 1 | 3 302 | 5.68 | 1 477 664 | 11.99 | 11.99 |
|  | 2 | 54 791 | 94.32 | 10 849 048 | 88.01 | 88.01 |
| PE8 | 1 | 5 946 | 10.24 | 1 636 128 | 13.27 | 13.27 |
|  | 2 | 52 147 | 89.76 | 10 690 584 | 86.73 | 86.73 |
| PE9 | 1 | 3 221 | 5.54 | 652 903 | 5.30 | 5.30 |
|  | 2 | 54 872 | 94.46 | 11 673 809 | 94.70 | 94.70 |
| SH1 | 1 | 1 153 | 1.98 | 330 993 | 2.69 | 2.69 |
|  | 2 | 56 940 | 98.02 | 11 995 719 | 97.31 | 97.31 |
| SH2 | 1 | 516 | 0.89 | 95 174 | 0.77 | 0.77 |
|  | 2 | 57 577 | 99.11 | 12 231 538 | 99.23 | 99.23 |
| SH3 | 1 | 1 081 | 1.86 | 361 067 | 2.93 | 2.93 |
|  | 2 | 57 012 | 98.14 | 11 965 645 | 97.07 | 97.07 |
| SH4 | 1 | 2 313 | 3.98 | 363 528 | 2.95 | 2.95 |
|  | 2 | 55 780 | 96.02 | 11 963 184 | 97.05 | 97.05 |
| SH5 | 1 | 129 | 0.22 | 53 252 | 0.43 | 0.43 |
|  | 2 | 57 964 | 99.78 | 12 273 460 | 99.57 | 99.57 |
| SH6 | 1 | 97 | 0.17 | 9 486 | 0.08 | 0.08 |
|  | 2 | 57 996 | 99.83 | 12 317 226 | 99.92 | 99.92 |
| SH7 | 1 | 3 246 | 5.59 | 359 771 | 2.92 | 2.92 |
|  | 2 | 54 847 | 94.41 | 11 966 941 | 97.08 | 97.08 |
| Total |  | 58 093 | 100.0 | 12 326 712 | 100.0 | 100.0 |

*For the labels of the disciplines see:
https://figshare.com/articles/dataset/OST_-_Classification_of_WoS_subject_categories_into_27_2_ERC_panels_/21707543

**A3. Influence plot (before excluding observations with disproportionate influence)**

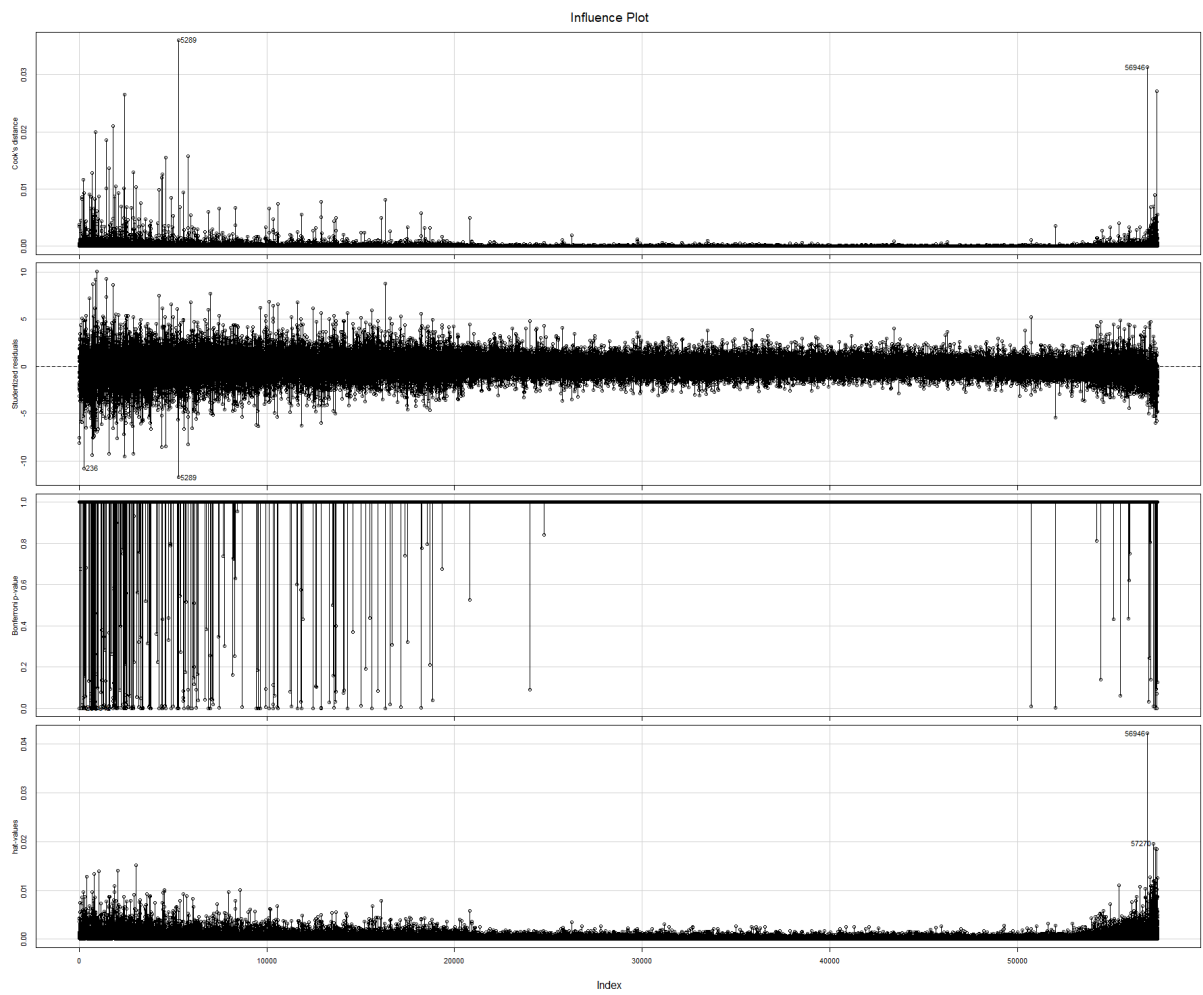

**A4. Model plots (before excluding observations with disproportionate influence)**

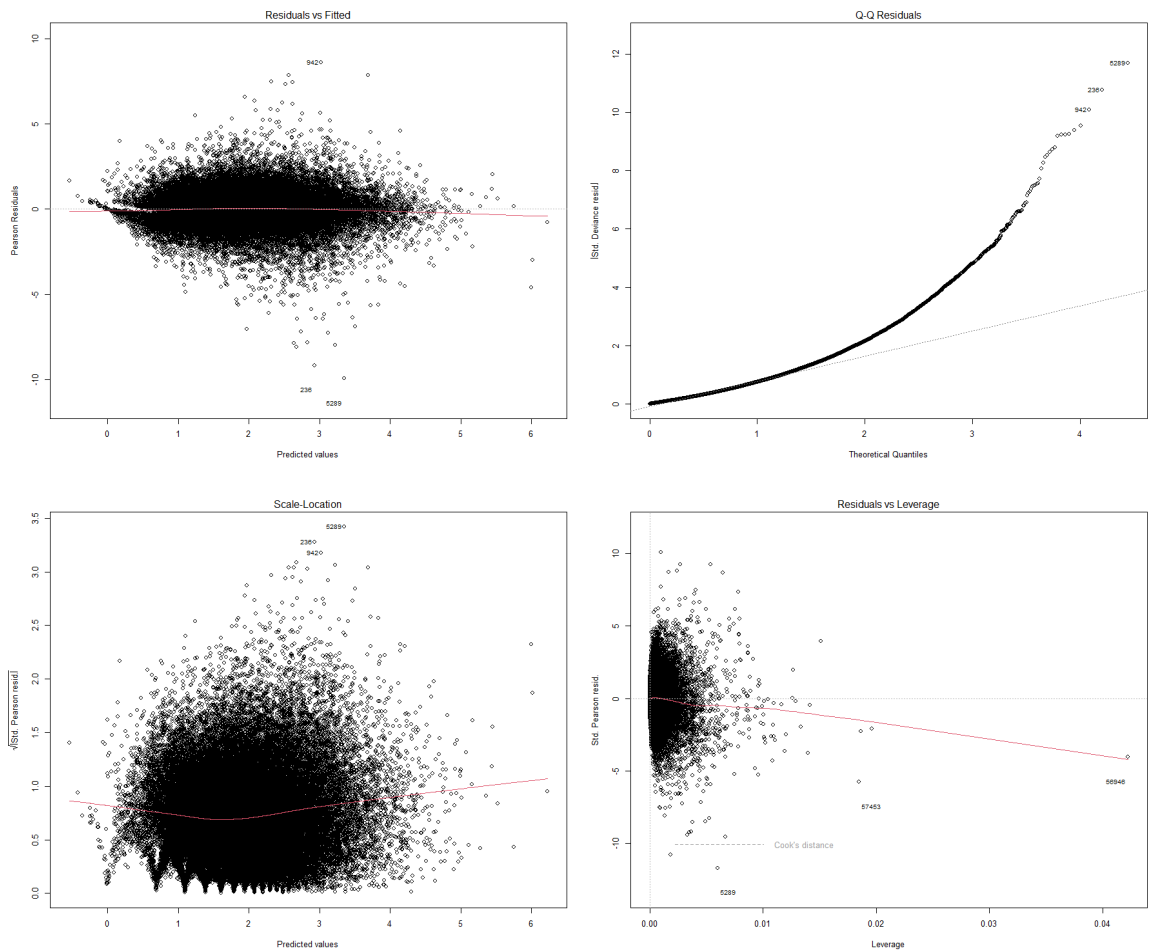

*Maddi, A., & Miotti, L. (2024). On the peer review reports: does size matter? Scientometrics. DOI: 10.1007/s11192-024-04977-6.*Maddi, A., & Sapinho, D. (2023). On the culture of open access: The Sci-hub paradox. *Scientometrics*, *128*(10), 5647–5658. https://doi.org/10.1007/s11192-023-04792-5

Moizer, P. (2009). Publishing in accounting journals: A fair game? *Accounting, Organizations and Society*, *34*(2), 285–304. https://doi.org/10.1016/j.aos.2008.08.003

Ni, P., & An, X. (2018). Relationship between international collaboration papers and their citations from an economic perspective. *Scientometrics*, *116*(2), 863–877. https://doi.org/10.1007/s11192-018-2784-9

Pautasso, M., & Schäfer, H. (2010). Peer review delay and selectivity in ecology journals. *Scientometrics*, *2*(84), 307–315. https://doi.org/10.1007/s11192-009-0105-z

Pranić, S. M., Malički, M., Marušić, S. L., Mehmani, B., & Marušić, A. (2020). Is the quality of reviews reflected in editors' and authors' satisfaction with peer review? A cross-sectional study in 12 journals across four research fields. *Learned Publishing*, *n/a*(n/a), 1–11. https://doi.org/10.1002/leap.1344

Price, D. J. D. S. (1963). Little Science, Big Science. In *Little Science, Big Science*. Columbia University Press. https://doi.org/10.7312/pric91844

Publons. (2018, February 26). *It's not the size that matters*. Publons. https://publons.com/blog/its-not-the-size-that-matters/

Quemener, J., Miotti, L., & Maddi, A. (2023). Technological Impact of Funded Research: A Case Study of Non-Patent References. *Quantitative Science Studies*.

Radzvilas, M., De Pretis, F., Peden, W., Tortoli, D., & Osimani, B. (2022). Incentives for Research Effort: An Evolutionary Model of Publication Markets with Double-Blind and Open Review. *Computational Economics*. https://doi.org/10.1007/s10614-022-10250-w

Shen, C., & Björk, B.-C. (2015). 'Predatory' open access: A longitudinal study of article volumes and market characteristics. *BMC Medicine*, *13*(1), 230. https://doi.org/10.1186/s12916-015-0469-2

Shen, S., Rousseau, R., Wang, D., Zhu, D., Liu, H., & Liu, R. (2015). Editorial delay and its relation to subsequent citations: The journals Nature, Science and Cell. *Scientometrics*, *105*(3), 1867–1873. https://doi.org/10.1007/s11192-015-1592-8
21